# Hybrid load balancing method with failover capability in server cluster using SDN

**Ok-Chol Ri[1], Yong-Jin Kim[2], and You-Jin Jong[1]**

[1] Kum Sung Middle School Number 2, PyongYang, 999093, D.P.R of Korea

[2] Faculty of Mathematics, KIM IL SUNG University, PyongYang, 999093, D.P.R of Korea

Corresponding author: Yong-Jin Kim (kyj0916@126.com).

This work was not funded at all.

**ABSTRACT** Traditional load balancers used in server clusters have problems such as lack of flexibility, high cost, etc. To overcome these problems, research has been conducted to apply a load balancer using software-defined network (SDN) to the server cluster. Under this trend, in this paper, we proposed a hybrid load balancing method with failover capability in the server cluster using SDN. The main goal of this paper is to improve the performance and failover capability of load balancing. First, we improved the performance of the dynamic weighted random selection (DWRS) algorithm by adopting the binary search to find the target server. Next, we proposed a dynamic method with failover capability by combining the improved DWRS algorithm with a fast-failover (FF) group table. In addition, a hybrid method combining the static method using the SELECT group table and the dynamic method with failover capability is proposed to improve the performance. Finally, the failover capability and performance of the proposed method are evaluated in an experimental environment.

## I. INTRODUCTION

Today, almost all Internet services are provided to users on a server cluster. Online services—such as search engines, websites, and social networks—are often replicated on multiple servers for greater capacity and better reliability [1]. Multiple servers get together to provide the same services, which can greatly increase computing capacity as well as reduce single points of failure, thus providing higher availability [2]. However, the massive number of servers and networking devices result in complexity in management, performance and scalability issues, and insufficient network bandwidth [3]. These problems are caused by load imbalance between server hosts and affect the performance of the server cluster. To solve this problem, a load balancing method is needed that can evenly distribute the client requests to the server hosts in the cluster.

Load balancing can be defined as the techniques employed to distribute client requests among an available set of servers with the objective of maximizing the throughput, minimizing the response time and avoiding the overload of the systems [4]. Load balancing is critical to the performance of a server cluster [5]. A suitable load balancing helps in maximizing scalability, minimizing response time, maximizing throughput, minimizing resource consumption, avoiding overload of any single resource and so on [6].

Traditional load balancing technology is often achieved by specific hardware that is usually very expensive and lacks sufficient flexibility [7]. Although it is fast, but is expensive and lack of flexibility in the configuration, making the configuration cannot be dynamically adjusted based on the network status [2]. To overcome these problems, research is being conducted to adopt a load balancer using the software-defined network to the server cluster.

SDN is an emerging architecture that decouples the control plane from the data plane and controls the network through a logically centralized controller (i.e., SDN controller), which allows better network management, control, and policy enforcement [8].

In traditional IP networks, the control plane and the data plane are both implemented inside the firmware of network devices such as routers and switches [5]. In such a network, if the forwarding policy is decided and implemented, it cannot be easily modified, and the modification can only be done by changing the characteristics of the entire network forwarding devices [9].

SDN is a state-of-the-art architectural approach to network management that allows more flexible management of a complicated large-scale network [5]. The task of the control plane is twofold: on the one hand, it is responsible for the administration of the switches, which will be instructed on the routing methods of the packets in transmission, on the other hand, it must serve to the higher level via creating an abstract and centralized vision of the underlying infrastructure [10]. The SDN controller is responsible for to definition and installation of these rules on the flow table of switches [11]. SDN promises flexibility, centralized control, and open interfaces between nodes, enabling an efficient, adaptive network [12]. This does not only reduce the need for manual configurations on switch, it can also provide greater flexibility in network management [7].

This feature of SDN enables the load balancer using SDN to overcome the problems of the traditional load balancer described above. According to our research, the performance of the previously published load balancing methods using SDN can be improved in many aspects. In this paper, we proposed a hybrid load balancing method with failover capability in the server cluster using SDN, and improved the performance and failover capability of load balancing.

The contributions of this paper are as follows.

1. We improved the performance of the dynamic weighted random selection algorithm by using binary search to find the target server.

2. We proposed a dynamic load balancing method with failover capability by combining the improved DWRS algorithm with the FF group table provided in OpenFlow.

3. A hybrid load balancing method combining the static method using the SELECT group table provided in OpenFlow with the dynamic method with failover capability is proposed to improve the performance.

4. Finally, we evaluated the failover capability and performance of the proposed method in the experimental environment.

## II. RELATED WORKS

Load balancing in a server cluster is divided into server load balancing and link load balancing [13]. Our study is a part of server load balancing. In this section, we have described the related works in the field of server load balancing.

According to our research, [14] is the first paper that introduced the load balancing method using the OpenFlow protocol [15]. In this paper, the authors use the net manager module and the host manager module to monitor and provide information such as network topology, network congestion, and server load value. The flow manager module uses this information to route flows and reduce response time.

Wang et al. [1] proposed a server load balancing method using OpenFlow wildcard rules. The authors presented an effective wildcard rule generation method for load balancing to save memory in switches. In addition, a method is provided to ensure that packets of the same TCP connection are handled on the same server hosts when the wildcard rule is changed. In this paper, a wildcard rule for load balancing is created statically without considering the load state of the server hosts.

Zhang et al. [16] presented a dynamic load balancing method that selects a target server based on the server load state determined by the number of client connections. In addition, the performance comparison result between the proposed method and the round-robin algorithm is provided. We believe that it is not ideal to determine the server load state according to the number of client connections. We believe that the server load state should be determined according to the overhead of the hardware resource.

Chen et al. [7] proposed a dynamic load balancing method that selects the target server for a client request according to the server load state. Unlike the above papers, in this paper, the authors calculated the server load value using CPU utilization, memory utilization, and network traffic values. They used the least-load algorithm, which selects the server host with the lowest load value as the target server. We believe that the method proposed here to determine the server load state is reasonable. The load monitoring method used in this paper can only be used in a virtual environment and cannot be used in a real server cluster environment.

Du et al. [2] determined the server load state based on the statistical traffic information captured by sFlow. Since the overhead of the server host for each client request is different, we believe that it is difficult to accurately evaluate the server load state using only the traffic statistical information.

Wang et al. [17] combine the static method, which uses a hash forward table, and the dynamic method, which uses the least load algorithm based on the load imbalance value of the server cluster. In this paper, the server load state is evaluated using the traffic statistical information provided in the flow table of the switch. As in the above paper, this load monitoring method cannot accurately provide the server load state.

In [18], the controller monitors the server load state using SNMP requests. In this paper, the least load algorithm is used to select the target server while considering the server health state.

In [19], the controller periodically sends a request to the server hosts and determines the load state based on the response time. The authors select the server host with the lowest response time as the target server.

TABLE 1 ANALYSIS RESULT OF RELATED WORKS

| Papers | Target server selection method | Failover capability |
|--------|-------------------------------|---------------------|
| [1] | Wildcard rule generation based on traffic statistics | No |
| [2] | Wildcard rule generation based on traffic statistics captured by sFlow | No |
| [5] | DWRS algorithm in a dedicated thread | No |
| [7] | Least-load algorithm | No |
| [8] | Least-load algorithm with request type consideration | No |
| [14] | Least-load algorithm | No |
| [16] | Least-connection algorithm | No |
| [17] | A hybrid method based on traffic statistics provided in the switch flow table | No |
| [18] | Least-load algorithm | Yes |
| [19] | Least-response time algorithm | No |
| [20] | Least-load algorithm in a dedicated thread | No |
| Our paper | A hybrid method based on the load imbalance value of the server cluster | Yes |

Abdelltif et al. [8] provided a dynamic load balancing method considering the request type. The authors classified the request type into computation request type and data request type, and proposed different server load value calculation methods for each type. The server load value is calculated using CPU utilization, memory utilization, and link bandwidth statistics, and the least load algorithm is used to select the target server.

In [20], the controller predetermines the target server in a dedicated thread. The authors reduce the overhead of selecting the target server and improve the overall system throughput by returning the predetermined target server immediately when the controller receives a packet-in message. The least-load algorithm is used to select the target server.

In [5], the controller predetermines the target server in a dedicated thread. [5] and [20] are both published by the same authors. The authors improved the system throughput more than in [20] by using the DWRS algorithm in the paper. We further improved the system performance by using the binary search algorithm [21] to find the target server after generating a random value in the DWRS algorithm.

Table 1 below shows our analysis results of related works.

## III. BACKGROUNDS
In this section, we introduce the traditional load balancing methods and OpenFlow group tables which are the important part of our proposed method.

### A. TRADITIONAL LOAD BALANCING METHODS
The results and discussion may be presented separately, or in one combined section, and may optionally be divided into headed subsections.

The load balancing method is divided into the static method and the dynamic method.

In the static method, the load balancing rule is generated based on the static information of the server cluster. Typical static methods include round-robin, random, weighted random selection (WRS), etc.

*Round-robin*: Server hosts in the cluster are sequentially selected as the target server for each request.

*Random*: A random server host in the cluster is selected as the target server for each request.

*WRS*: In this method, requests are probabilistically distributed among the server hosts based on the predetermined server capacity values.

In the dynamic method, the load balancing rule is generated based on the current load state of each server host. Typical dynamic methods include least-load, least-connections, least-response time, DWRS, etc.

*Least-load*: In this method, the load balancer periodically monitors the hardware resource overhead of each server host and determines the load value. The server host with the least load is selected as the target server for the request.

*Least-connections*: In this method, the load balancer periodically monitors the number of client connections of each server host. The server host with the least number of client connections is selected as the target server for the request.

*Least-response time*: In this method, the load balancer periodically monitors the response time of each server host. The server host with the least response time is selected as the target server for the request.

*DWRS*: In this method, the load balancer periodically monitors the capacity value of each server host in the cluster. Requests are probabilistically distributed to the server hosts based on these capacity values. The load balancer first generates a random value that is smaller than the sum of the capacity values of the server hosts to select the target server. Next, find the smallest n where the sum of the capacity values of the first n server hosts is greater than this random value, and this n-th server host is selected as the target server. In this method, the linear search is used to find this n. In our study, we reduced the overhead for finding n by adopting the binary search in this step.

### B. OPENFLOW GROUP TABLES
In addition to the flow table, OpenFlow provides group tables for effective flow control.

A group table consists of multiple group entries, and each entry contains four fields: group identifier, counters,

action buckets, and group type. The meaning of each field is described below.

*Group identifier*: a 32bit unsigned integer used for uniquely identifying each group table.

*Counter*: updated when the packets are processed by the group table.

*Action buckets*: An ordered list of multiple bucket entries. Each bucket entry contains an action set, the bucket weight used in the SELECT group table, and the watch group/port used in the SELECT/FF group table.

*Group type*: can be one of the four values: ALL, SELECT, INDIRECT, and FF.

The function of each group type is as follows.

*ALL*: The packet is forwarded to all buckets of the group table simultaneously. After the packet arrives, the action set in each bucket is applied to the packet. This type of group table can be used for packet multicast or broadcast.

*INDIRECT*: Supports only a single bucket and can be seen as an ALL group table with a single bucket.

*SELECT*: When a packet arrives, the group table selects a bucket based on a switch-computed selection algorithm. The packet is sent to the selected bucket and the action set in this bucket is applied to the packet. Switch-computed selection algorithms include several algorithms such as round-robin, hash-based selection, and so on. In the round-robin algorithm, buckets in the group table are sequentially selected for packet processing. In the hash-based selection algorithm, the bucket is selected according to the hash result of the hash-field data in the packet. Fields such as IP/MAC address of source/destination host, VLAN tags, and Ether type can be included in the hash-field. If the bucket weight is specified for each bucket entry, the group table probabilistically selects the bucket based on this value. If no selection method is specified, Open vSwitch up to release 2.9 applies the hash method with default fields [22]. The hash-based selection guarantees that all packets belonging to the same TCP session are forwarded via the same interface [23]. When a port specified in a bucket in a SELECT group goes down, the switch may restrict bucket selection to the remaining set (those with forwarding actions to live ports) instead of dropping packets destined for that port [15].

*FF*: When a packet arrives, the action set of the first live bucket in the group table is applied to the packet. In this type of group table, the watch port/group field is set for each bucket. If the server host connected to the bucket's watch port is live or if the watch group set in the bucket exists, that bucket is called a live bucket. This group table can be used to provide an effective failover capability in a server cluster. We have combined this group table with the load balancing method to provide the failover capability in load balancing.

## IV. HYBRID METHOD WITH FAILOVER CAPABILITY

In this section, we proposed a hybrid load balancing method with failover capability in the server cluster using SDN. We first described the load monitoring module that initializes the cumulative sum list of the serviceability values of each server host. Next, the static load balancing method using the hash-based selection algorithm of the SELECT group table is described. Next, the improved DWRS algorithm that finds the target server by applying binary search to the cumulative sum list initialized in the load monitoring module is described. In addition, the dynamic method with failover capability combining the improved DWRS algorithm and the FF group table is also described. Finally, a hybrid load balancing method that combines the static method and the dynamic method based to the load imbalance value of the server cluster is described.

### A. LOAD MONITORING MODULE

This module periodically monitors the load values of server hosts in the cluster and initializes the cumulative sum list of serviceability values. The serviceability is the ability of each server host to handle the requests, and the maximum of the serviceability value is 1. The serviceability value is defined as the value obtained by subtracting the load value from 1. Therefore, the higher the server load value, the lower the serviceability value. The cumulative sum list is used to find the index of the target server in the improved DWRS algorithm. The n-th element of the list is equal to the sum of the serviceability values of the first n server hosts, and therefore this list is a monotonically increasing list. The value of each element of the cumulative sum list is defined as follows.

$$s_i = \begin{cases} 1.0 - x_0 & , \ i = 0 \\ s_{i-1} + (1.0 - x_i), & 0 < i < n \end{cases} \quad (1)$$

In the above equation, $s_i$ is the value of the i-th element of the cumulative sum list, $x_i$ is the load value of the i-th server host, and n is the size of the server cluster.

In the improved DWRS algorithm that will be described below, the index of the target server can be accurately found only when no server failure occurs. For example, when a failure occurs in the 3rd and 6th server hosts in a cluster of size 8, the controller only receives the load values of the other 6 live server hosts. The index of the target server found by applying the improved DWRS algorithm to these 6 server hosts cannot be the same as the index in the cluster. Therefore, to make this algorithm work even in the case of server failure, we initialize the cumulative sum list in a specific way. We set the load value of the failed server host to 1.0 and initialize the cumulative sum list. Table 2 below shows the initialization results of the serviceability values and the cumulative sum list, when a failure occurs in the 3rd and 6th server hosts and the load values of 6 other live server hosts are given.

| index | 0 | 1 | 2 | 3 | 4 | 5 | 6 | 7 |
|---|---|---|---|---|---|---|---|---|
| live state of the i-th server host | yes | yes | yes | no | yes | yes | no | yes |
| load value of the i-th server host | 0.1 | 0.4 | 0.3 | 1.0 | 0.7 | 0.2 | 1.0 | 0.5 |
| serviceability value of the i-th server host | 0.9 | 0.6 | 0.7 | 0.0 | 0.3 | 0.8 | 0.0 | 0.5 |
| the i-th element of the cumulative sum list | 0.9 | 1.5 | 2.2 | 2.2 | 2.5 | 3.3 | 3.3 | 3.8 |

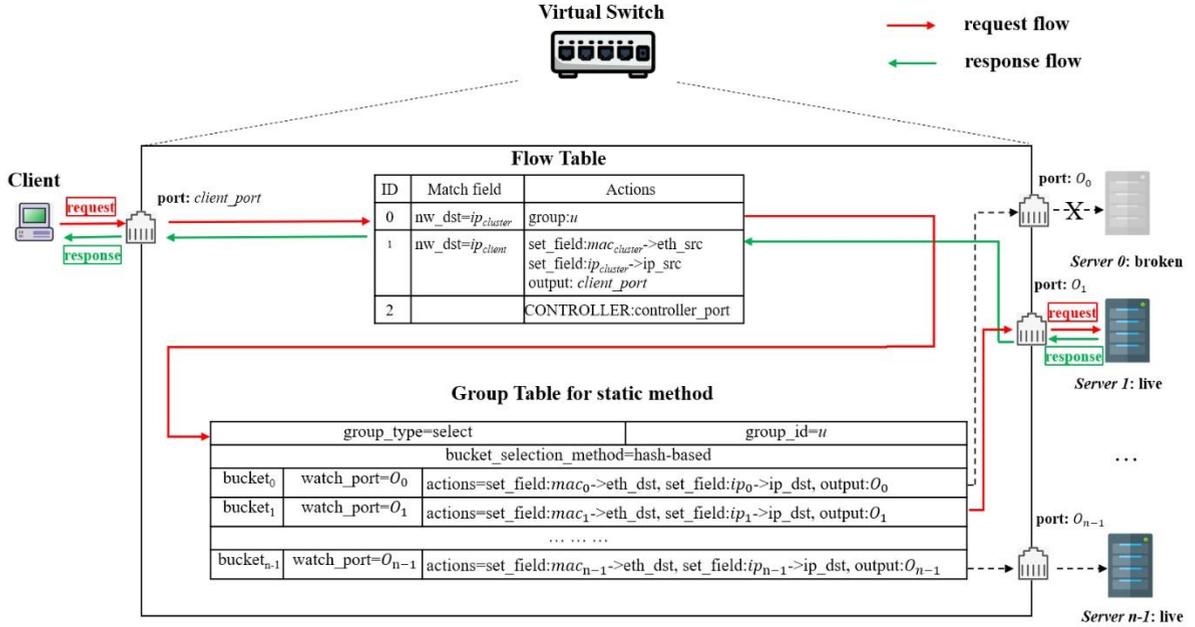

**FIGURE 1. Packet handling process in the static load balancing method.**

Since a broken server cannot handle any requests, we believe it is reasonable to set its load value to 1.0. When the improved DWRS algorithm is applied to this cumulative sum list, other live server hosts than the 3rd and 6th are selected as target servers. For example, in the above case, if the random value to be found is 2.2, the enhanced DWRS algorithm selects the 2nd server host as the target server. This is because the improved DWRS algorithm selects the first element if there are multiple elements in the list that are equal to the value to be found.

### B. STATIC METHOD USING GROUP TABLE

We proposed a static method using the SELECT group table provided in OpenFlow. This method is a switch-based method that selects the target server by using the hash-based selection algorithm of the SELECT group table without the participation of the controller. Therefore, the performance of load balancing can be further improved. According to our experiments, the load balancing throughput of our static method is 2.42 times that of the round-robin algorithm.

The design of our static method is as follows.

First, we inserted a flow entry into the switch's flow table, which is applied to the request packets sent from the client to the server cluster. The flow entry applies a specific action set to the packets whose destination IP address is the virtual IP address (VIP) of the server cluster. This action set forwards the packet to the SELECT group table.

Next, we insert a SELECT group table into the switch and insert multiple buckets into this group table. The number of buckets is equal to the size of the server cluster. Each bucket is associated with a different server host in the cluster and contains a set of actions that will forward packets to the associated server host. For each bucket, we set the watch port as the switch port number that connects the associated server host to the switch. This group table distributes client requests to server hosts, considering server failures. The SELECT group table provides the hash-based selection algorithm and the round-robin algorithm for bucket selection. We use the hash-based selection algorithm to avoid handling requests from the same client on different target servers.

In addition, a flow entry applied to the response packets sent from the server cluster to the client is also inserted into the switch's flow table. The flow entry applies a specific action set to the packets whose destination IP address is the client's IP address. This action set modifies the source IP address field of the packet to the VIP of the server cluster.

The packet handling process in the static method is shown in Fig.1.

In the figure, *client_port* is the switch port number connected with the client, *u* is the group identifier of the SELECT group table, $ip_{cluster}$ is the VIP of the server cluster, $ip_{client}$ is the IP address of the client, $mac_{cluster}$ is the MAC address of the server cluster, $mac_i$ is the MAC address of the i-th server host, $ip_i$ is the IP address of the i-th server host, $O_i$ is the switch port number connected to the i-th server host. As shown in the figure, if the server host associated with $bucket_0$ is down, the bucket to handle the request is selected from among other buckets.

### C. DYNAMIC METHOD WITH FAILOVER CAPABILITY

In this section, we proposed a dynamic load balancing method with failover capability by combining the improved DWRS algorithm and the FF group table provided in OpenFlow. First, we proposed an improved DWRS algorithm that uses binary search to find a target server. Next, we describe the failover capability provided by using the FF group table.

#### 1) MOTIVATION TO SELECT DWRS

The dynamic methods we have explored include least-load, least-connections, and DWRS.

The least-connections algorithm determines the server host load value based on the number of client connections. The least-load algorithm determines the load value according to the hardware resource overhead of the server host. The least-load algorithm can monitor the load state of server hosts more accurately than the least-connections algorithm, and therefore we believe that the least-connections algorithm does not have higher performance than the least-load algorithm.

We also believe that the least-load algorithm has a lower performance compared to DWRS. The reason is as follows.

The goal of load balancing is to minimize the load imbalance value and maximize the throughput in the server cluster. In the least-load algorithm, when a request is received, the request is sent to the previous target server selected at the previous monitoring time. Therefore, all requests received within a monitor interval are aggregated to the previous target server. In the real environment, many requests are received within a single monitoring interval. In this case, the load value of the previous target server increases rapidly at the next monitor time. Therefore, the load imbalance value of the server cluster cannot keep the optimal value, and the load balancing performance is affected. If the monitoring interval is shortened to solve this problem, the monitoring overhead of the controller increases, which also affects the load balancing performance.

However, in the DWRS algorithm, all requests received within a monitoring interval are not sent only to the previous target server. This algorithm distributes the request to the server hosts probabilistically based on the load value. Therefore, this algorithm can prevent the rapid increase of the load value of a certain server host and promote the performance improvement. According to our experiment, the load balancing throughput of the DWRS algorithm is 1.29 times that of the least-load algorithm. The comparison result between the DWRS algorithm and the least-load algorithm is also provided in [5].

#### 2) IMPROVED DWRS USING BINARY SEARCH

In the DWRS algorithm, a random value smaller than the sum of the serviceability values of the server hosts is generated, and the first element of the cumulative sum list greater than this random value is found using linear search, and that element is selected as the target server. We reduce the overhead of this step by using binary search instead of linear search. The binary search algorithm has a complexity of $O(log(n))$, which is much more powerful than a linear search algorithm whose complexity is $O(n)$.

Algorithm 1 is the corresponding pseudo-code.

ALGORITHM 1: TARGET SERVER SELECTION USING BINARY SEARCH IN DWRS

*n: the size of the server cluster*
*ts: selected target server*
*s[1…n]: cumulative sum list of serviceability values*
*r: generated random value*
*lw_bound=1, up_bound=n: lower and upper boundaries for binary search*
**While** *lw_bound + 1 < up_bound:*
        *mid = (lw_bound + up_bound) / 2*
        **If** *(r <= s[mid])* **then**
                *up_bound = mid;*
        **Else**
                *lw_bound = mid;*
        **End if**
**End while**
*ts = up_bound;*
return *ts;*

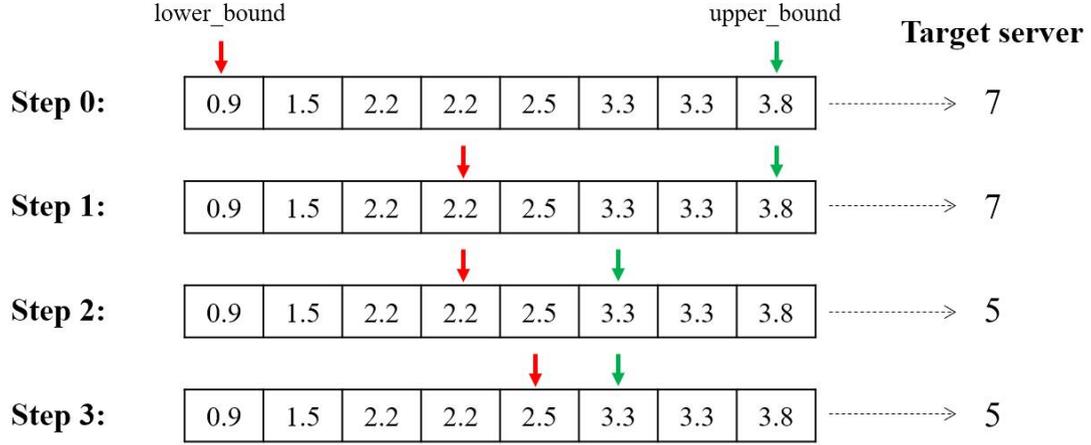

**FIGURE 2.** Target server selection steps using binary search.

Fig.2 shows the target server selection steps using the binary search when the random value to be found is 3.3.

As shown in the figure, the binary search requires 3 steps to find the target server, but the linear search requires 6 steps.

Of course, in this method, the overhead for initializing the cumulative sum list cannot be ignored. We initialize this list at the time of load monitoring. After the list is initialized, for all packet-in messages received until the next load monitoring time, the binary search is applied to the already initialized list to find the target server. However, in the traditional DWRS, the linear search is applied for all packet-in messages received until the next load monitoring time, therefore, the improved DWRS can further improve the performance compared to the traditional DWRS.

After the target server is selected, the controller inserts a flow entry that applies a specific action set to the request packet. This action set generally modifies the destination IP address field of the packet to the IP address of the target server and outputs the packet. However, we have modified this action set to provide the failover capability. The modified action set forwards the packet to the FF group table associated with the target server. The FF group table used here is described in detail in the next section.

### 3) FAILOVER CAPABILITY USING GROUP TABLE
We provide real-time failover capability using the FF group table. We have defined a associated FF group table for each server host in the cluster.

#### a: FF GROUP TABLE CONFIGURATION
Each FF group table consists of multiple buckets to provide failover when a failure occurs in a server host associated with it. Each bucket is associated with a different server host. The first bucket is associated with the target server associated with the group table and the other buckets are associated with the target server's standby servers. The closer the load value of the standby server to the target server, the smaller the index of the bucket associated with that standby server. Each bucket applies a specific action set to the request packet. This action set forwards the packet to the server host associated with its bucket. In addition to the forwarding action set, each bucket has a watch port associated with it. The watch port field saves the switch port number connected to the server host associated with its bucket.

*How many standby servers should be selected?* The number of standby servers depends on the design of the server cluster. We have determined p-1 as the number of standby servers when the probability of simultaneous failure of p or more server hosts in the cluster is almost 0.

*How the standby servers should be selected?* We select the standby servers of each server host so that the load imbalance value of the cluster can keep the minimum value even after the failover. We select p server hosts with the closest load value as standby servers for each server host.

#### b: WHEN SHOULD THE FF GROUP TABLE BE UPDATED?
We update the FF group table when the load state of the server hosts changes, including the case of a server host failure. In other words, the FF group table is updated when there is a server host whose standby servers are changed to other ones. We have defined a specific threshold for detecting changes in the load state on the server host. If there is a server host whose load change value is greater than this threshold, the FF group table will be updated.

*How this threshold should be set?* Even when the server host is not handling any client requests, CPU utilization fluctuates with small amplitudes. We considered that the load state changed when the load change value of server hosts is larger than this amplitude value. Let L = $\{l_0, l_1, \dots l_m\}$ ($0 \leq l_i \leq 1$) be the set of CPU utilization measured on a server host during a certain interval under a no-load condition. We define the threshold as follows.

$$t = \max_{0 \leq i \leq |L|} l_i - \min_{0 \leq i \leq |L|} l_i \quad (2)$$

The threshold we have defined is the amplitude of CPU utilization of a server host under a no-load condition. This value is very small, therefore, when a failure occurs in the server host and the load value becomes 1.0, the load change value is greater than this threshold. Finally, the FF group table is updated when the load state of the server hosts changes, including when a failure occurs in the server host.

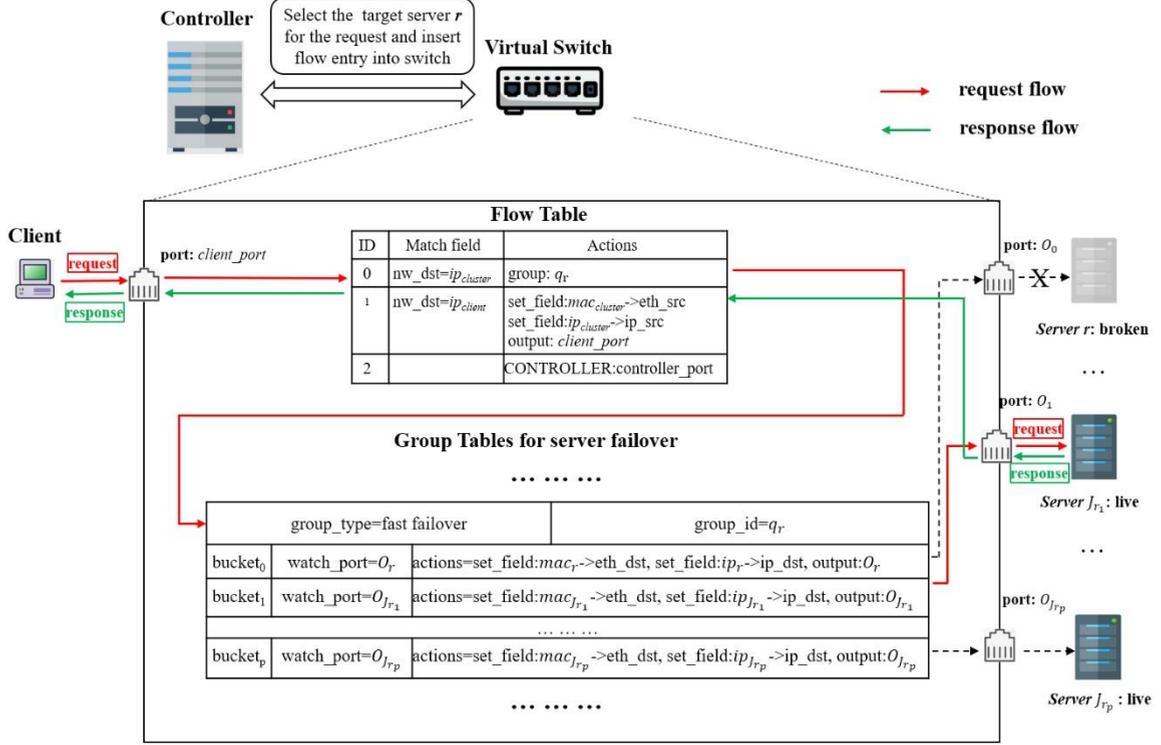

**FIGURE 3.** The working flow of failover mechanism using FF group table.

## C: HOW TO COMBINE THE GROUP TABLE WITH LOAD BALANCING?

When the target server is selected, the controller generally inserts a flow entry with a specific action set into the switch's flow table. This action set sends the request packet to the selected target server. The action set modifies the IP/MAC address of the destination host in the request packet to the address of the target server and outputs the packet to the switch port connected to the target server. However, in the proposed method, we set this action set to forward the packets to the FF group table associated with the selected target server. The FF group table applies the action set of the first live bucket when it receives the packet. For example, if the target server associated with the first bucket is already down, the action set of the second bucket is applied and the packet is forwarded to the appropriate standby server. Therefore, even if the target server fails, the packet is not lost and is handled by the cluster. The working flow of the failover mechanism using the FF group table is shown in Fig.3.

In the figure, $r$ is the index of the selected target server, $p$ is the number of standby servers, $q_r$ is the group identifier of the FF group table associated with the r-th server host, and $J_{r_i}$ is the index of the i-th standby server of the r-th server host. The meanings of other symbols are the same as in Fig.1.

### D. HYBRID METHOD

In this section, we propose a hybrid load balancing method that combines the static and dynamic methods discussed above.

#### 1) MOTIVATION FOR THE HYBRID METHOD

The dynamic method, which provides load balancing based on server load values, may not be the best in all conditions. For example, if the load values of the server hosts are almost the equal, the switch does not need to communicate with the controller and it can select the best target server even with a static method. However, in the dynamic method, the switch must communicate with the controller to select the target server after receiving the packet. Therefore, in this case, the response may be delayed due to this unnecessary communication overhead, which may affect the load balancing performance. According to our experiments, the load balancing throughput of the static method, in this case, is 2.31 times that of the dynamic method. Therefore, we proposed a hybrid method that combines the static and dynamic methods.

## 2) DESIGN OF THE HYBRID METHOD

We switch between the static method and dynamic method based on the load imbalance value of the server cluster, and this value is monitored in real-time. First, the cluster administrator needs to set a certain threshold. Next, we activate the static method when the load imbalance value of the server cluster is less than this threshold, and activate the dynamic method in the other case.

We have defined the variance of server load values as the load imbalance value of the cluster.

$$\delta = \frac{\sum_{i=0}^{n-1}(x_i - \bar{x})^2}{n} \quad (3)$$

In the above equation, $x_i$ is the load value of the i-th server host, $\bar{x}$ is the average load value of the server hosts, and n is the size of the server cluster.

When switching from the dynamic method to the static method, the controller first deletes the flow entries inserted in the dynamic method. We delete flow entries where the destination IP address field of the match field is the IP address of the server cluster or client. Next, we insert the SELECT group table and the flow entries which are used for static load balancing. These flow entries are applied to the request and response packets. The function of this group table and the flow entries is discussed in detail in Section 4.2.

When switching from the static method to the dynamic method, the controller deletes the SELECT group table and flow entries that were inserted in the static method. After that, requests are handled based on the improved DWRS algorithm installed in the controller.

## V. EVALUATION

In this section, we evaluated the failover capability and performance of the proposed method in the experimental environment. First, we described the environment configuration for the experiment. Then, we evaluated the failover capability by measuring the packet loss rate. We also provide the comparison results for throughput and load imbalance value metrics between round-robin, least-load, DWRS algorithms, and the proposed method.

## 1) ENVIRONMENT CONFIGURATION

We need a controller, a switch, a client, and a cluster of multiple server hosts for the evaluation. According to our research, mininet [24] is a powerful tool for simulating a software-defined network environment consisting of multiple controllers and switches and thousands of client hosts on a single physical host. In many of the previous works in the literature we reviewed, mininet is used to configure the experimental environment. Our dynamic load balancing method using the binary search has better performance advantages as the size of the server cluster increases. Therefore, mininet, which can simulate thousands of virtual hosts, can provide an ideal experimental environment. However, we do not use mininet because it cannot satisfy important requirements for the performance evaluation. The reason is as follows.

In the dynamic method, we select the target server based on the CPU utilization of the server hosts in the cluster. Therefore, the simulated server hosts need to work on independent CPU resources. According to our experiment, the simulated hosts in mininet cannot satisfy this requirement, because they share the CPU resource of the physical host. In other words, the CPU utilization of each server host in the cluster always changes uniformly. Therefore, we configure the environment with virtual machines running on VMWare and Virtual Box without using mininet.

The topology of the experimental environment is shown in Fig.4.

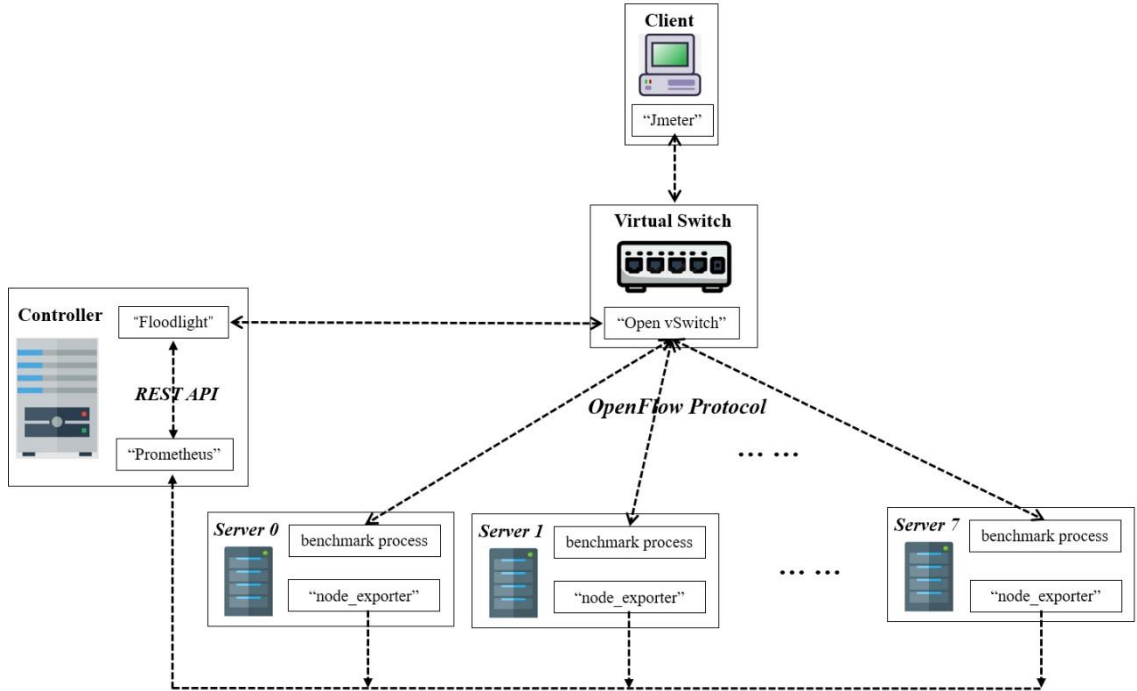

**FIGURE 4. Environment configuration for evaluation**

TABLE 3: THE CONFIGURATION FOR EACH VIRTUAL HOST IN THE EXPERIMENTAL ENVIRONMENT.

|  | Controller | Switch | Client | Server host$_{0-7}$ |
|---|---|---|---|---|
| OS Version | Ubuntu 14.04.06 desktop | Ubuntu 14.04.06 desktop | Windows 10 Enterprise | Ubuntu 14.04.06 server |
| Installed Software | Floodlight and Prometheus 2.36.0 | Open vSwitch 2.7.0 | Jmeter 5.5 | Benchmark process and Node_exporter 1.3.1 |
| IP address | 192.168.198.130 | 192.168.198.150 | 192.168.198.120 | 192.168.198.151-158 |
| Hardware configuration | 2 CPUs, 4GB RAM | 1 CPU, 2GB RAM | 1 CPU, 2GB RAM | 1 CPU, 512MB RAM |

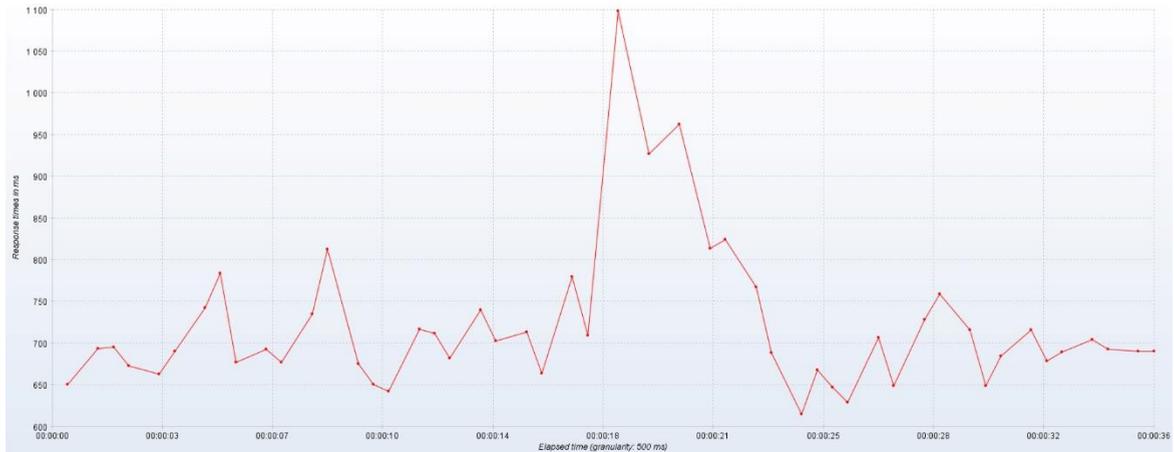

**FIGURE 5. The changing state of response time when server failure occurs in the proposed method**

We simulated the virtual hosts using VMWare and Virtual Box on the physical host with Intel(R) Core™ i7-8750H CPU@2.2GHz (12CPUs) and 16GB of RAM. The configuration for each virtual host is shown in Table 3.

*Why Prometheus?* We use Prometheus [25] to monitor the load state of server hosts every 10 seconds. We executed the node_exporter process provided in Prometheus on each server host. This process periodically sends the load state of the server hosts to the Prometheus process running on the controller. Node_exporter can monitor many metrics for the server host, including CPU, memory, and I/O load values. These metrics are calculated on the server host and sent to the controller, so the controller overhead for monitoring is not high. The Prometheus process provides real-time server host load state in a graphic interface mode. This allows the user to visually monitor the load balancing status of the server cluster. Less controller overhead for monitoring and providing visual monitoring results are the reasons why we use Prometheus for monitoring.

In the controller, the Prometheus and the Floodlight processes run together. Floodlight process uses HTTP API to query the load value of each server host monitored by the Prometheus process. The monitoring results are used to select the target server in Floodlight process.

*How to configure Jmeter?* According to our measurement, the most ideal number of users to evaluate the throughput of the proposed method in our environment is 40. Therefore, we set the thread count in Jmeter to 40. This value varies depending on the overhead of the benchmark code and the experimental environment.

We run a benchmark process that calculates certain mathematical formulas on the server host for evaluation, and this process is written in Python.

We evaluated the failover capability by measuring the packet loss rate. We shut down one server host in the cluster after 19 seconds from when the client started sending the request and analyzed the changes in packet loss rate and response time. Fig.5 shows the changing state of the response time that we measured.

As shown in Fig.5, the response time increased rapidly to 1100ms after 19 seconds from the start of the experiment. The switch detects that a failure has occurred in the target server selected by the controller and sends the request to the appropriate standby server. As a result, the response time was delayed, but no packet loss occurred.

We also evaluated the failover capacity of the round-robin, which is one of the default load balancing methods provided in Floodlight. The failover capacity is not provided in this algorithm. Fig.6 shows the changing state of response time in the round-robin algorithm when server failure occurs.

## 2) FAILOVER CAPABILITY

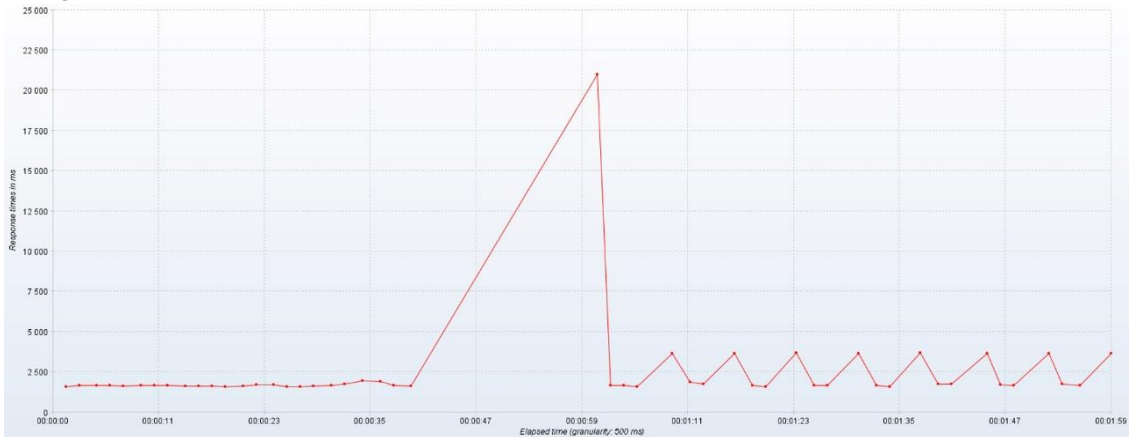

**FIGURE 6. The changing state of response time when server failure occurs in round-robin**

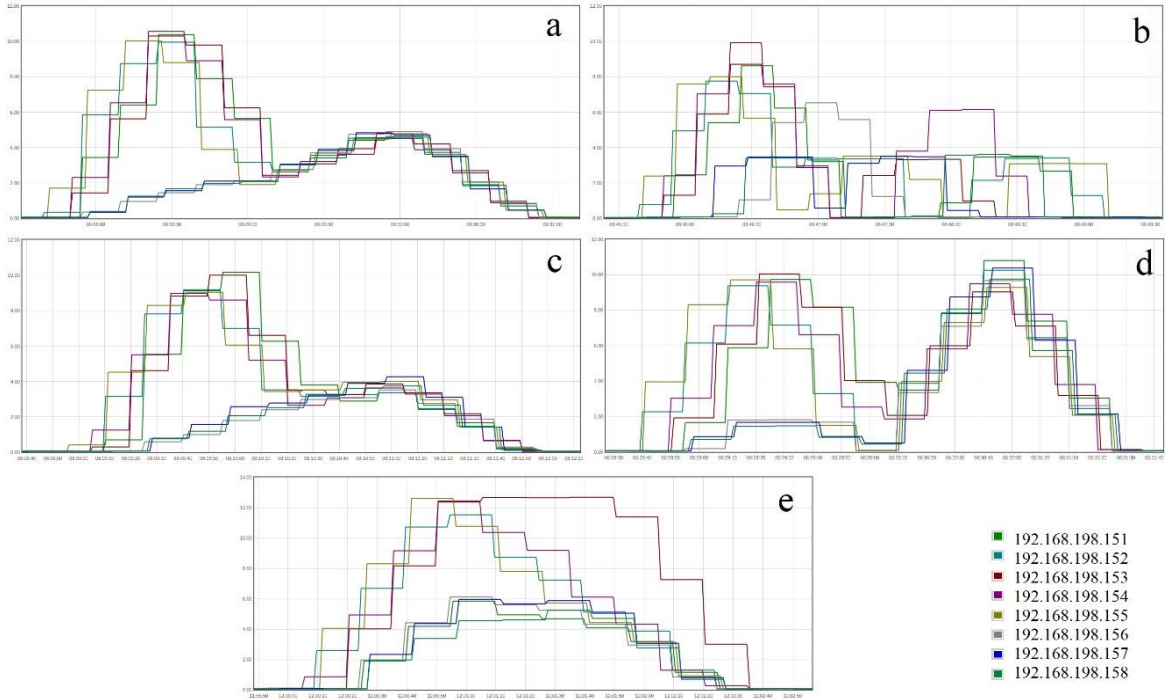

**FIGURE 7. Load state change graph of server hosts in cluster monitored by Prometheus. (a) Round-robin. (b) Least-load. (c) DWRS. (d) Proposed method (threshold=0.005). (e) Proposed method (threshold=0.01).**

In this algorithm, the packet loss has occurred and the loss rate is 2%. After 60 seconds from the start of the experiment, the response time increased rapidly to 20 seconds. Before the experiment, we set the connection timeout parameter in Jmeter to 20 seconds. If the response time exceeds this value, it means that the request packet is lost due to server failure.

### 3) PERFORMANCE

We provide the comparison results for throughput and load imbalance value metrics between round-robin, least-load, DWRS algorithms, and the proposed method. Since there is an offset in the experimental environment for each experiment, we measured 3 times and recorded the average value as the real value. To simulate the real network environment, we set different load values of server hosts before the experiment. We run the process of calculating the value of $\pi$ to 5000 decimal places by using the "bc" command of Linux on the first 5 server hosts in the cluster. According to our measurement, after this process is executed, the CPU utilization of the server host increases to about 13%, keeps this value for 1 minute, and then returns to the initial state.

Our experimental results for round-robin, least-load, DWRS, and the proposed method are shown in Fig.7 and Table 4.

TABLE 4: EXPERIMENTAL RESULTS FOR ROUND-ROBIN, LEAST-LOAD, DWRS, AND THE PROPOSED METHOD

|  | Throughput [requests/s] | Load imbalance |
|---|---|---|
| Proposed method (threshold=0.01) | 26.926810 | 0.003995 |
| Proposed method (threshold=0.005) | 18.600817 | 0.003038 |
| Round-robin | 16.276773 | 0.003172 |
| Least-load | 13.231500 | 0.003104 |
| DWRS | 16.133330 | 0.017145 |

According to the experimental results, the performance of the least-load is worse than the round-robin. This reason is explained above.

In the experiment, we set the threshold of the proposed method to 0.5% and 1%, respectively. This value is set by the administrator according to the state of the server hosts in the cluster. As shown in Table 4, when the threshold is set to 0.5%, the throughput is reduced by 30.8% and the load imbalance value is improved by 23% compared to when the threshold is set to 1%. When the threshold is set small, the dynamic method is activated for most requests. In this case, the communication overhead between the switch and the controller increases and the throughput decreases. Conversely, if the threshold is set high, the static method is used for most requests. In this case, because the target server is selected without regard to the state of the server hosts, the load imbalance value increases, and the throughput may also decrease. Therefore, it is important to set the threshold to the right value.

Experimental results are measured values in our experimental environment, so they do not have a reference value. They were used only to compare the performance of the proposed method with round-robin, least-load, and DWRS.

## VI. CONCLUSION

In this paper, we have proposed a hybrid load balancing method with failover capability in the server cluster using SDN. The main goal of this paper is to improve the performance and failover capability of load balancing. We proposed a hybrid load balancing method that combines the static method using the SELECT group table and the dynamic method with failover capability. We evaluated the failover capability by measuring the packet loss rate and provided the performance comparison results between the proposed method and traditional load balancing methods in the experimental environment.

It is important to set the load imbalance threshold to the right value in the proposed method. In our experiment, when the threshold is set to 0.5%, the throughput is reduced by 30.8% and the load imbalance value is improved by 23% compared to when the threshold is set to 1%. The cluster administrator can adjust the threshold according to the priority of each metric.

The improved DWRS algorithm using binary search shows better performance advantages compared to the traditional DWRS algorithm as the size of the server cluster increases. As a hardware limitation, we evaluated the proposed method on a cluster consisting of only 8 server hosts. In the future, we will configure an experimental environment more similar to the real environment with enough server hosts and provide the performance evaluation results. We will also investigate not only server load balancing but also link load balancing in the server cluster.

## ACKNOWLEDGMENT
The source code of the hybrid load balancing controller used to support the results of this study is available at https://github.com/r-o-c-2022/sdn_controller_with_hybrid_load_balancer.

The authors declare that there is no conflict of interest regarding the publication of this paper.

We thank the anonymous reviewers, the Associate Editor, and the Editor-in-Chief for their valuable feedback on the paper.